\documentclass[twocolumn,aps,superscriptaddress]{revtex4-2}

\usepackage{amsmath, amssymb, amsthm}
\usepackage{amsfonts}
\usepackage{graphicx}
\usepackage{epsfig}   
\usepackage{subcaption}

\usepackage{tabularx}

\usepackage{xcolor}
\usepackage[normalem]{ulem}   

\usepackage{comment}

\renewcommand{\sout}{\bgroup \color{red} \ULdepth=-.5ex \ULset}

\usepackage[font=small,labelfont=bf,textfont=it]{caption}

\begin{document}

\title{Exploring the chiral magnetic effect in Au+Au collisions at $\sqrt{s_{NN}}=7.7-200$ GeV through Chiral Anomaly Transport}

\author{Zilin Yuan}
\email{yuanzilin20@mails.ucas.ac.cn}
 \affiliation{School of Nuclear Science and Technology, University of Chinese Academy of Sciences, Beijing, 100049,  P.R. China}
\affiliation{Institue of High Energy Physics, Chinese Academy of Sciences,
Beijing, 100049,  P.R. China}
\author{Anping Huang}
\email{huanganping@ucas.ac.cn}
\affiliation{School of Nuclear Science and Technology, University of Chinese Academy of Sciences, Beijing, 100049,  P.R. China}
\affiliation{School of Material Science and Physics, China University of Mining and Technology, Xuzhou, China}
\author{Guo-Liang Ma}
\email{glma@fudan.edu.cn}
\affiliation{Key Laboratory of Nuclear Physics and Ion-beam Application (MOE), Institute of Modern Physics, Fudan University, Shanghai 200433, China}
\affiliation{Shanghai Research Center for Theoretical Nuclear Physics, NSFC and Fudan University, Shanghai $200438$, China}
\author{Mei Huang}
\email{huangmei@ucas.ac.cn}
\affiliation{School of Nuclear Science and Technology, University of Chinese Academy of Sciences, Beijing, 100049,
  P.R. China}
\author{Guannan Xie}
\email{xieguannan@ucas.ac.cn}
\affiliation{School of Nuclear Science and Technology, University of Chinese Academy of Sciences, Beijing, 100049,  P.R. China}
  
\begin{abstract}
High-energy heavy-ion collisions have the potential to create local domains of chirality-imbalanced quarks, reflecting the topological characteristics of quantum chromodynamics. This phenomenon can potentially induce local $\mathcal{P}$ and $\mathcal{CP}$ violations in the quark-gluon plasma. The Chiral Magnetic Effect (CME) predicts an electric charge separation along the intense magnetic field generated during these collisions, which is typically investigated through charge-dependent azimuthal correlations ($\Delta\gamma$). In this work, we investigate the CME in Au+Au collisions at $\sqrt{s_{NN}} = 7.7 - 200$ GeV using a multiphase transport (AMPT) model equipped with a Chiral Anomaly Transport (CAT) module. we employ two independent methods: direct subtraction of the correlator $\langle N_{part}\Delta\gamma\rangle$ between simulations with zero and finite chiral chemical potential $\mu_5$, and the event-shape-selection (ESS) approach.
Our results reveal a significant CME signal within the energy range of $11.5-27$ GeV and the centrality range of $20-50\%$, where the AMPT model aligns well with STAR experimental data. Furthermore, the CME fractions extracted by both methods are consistent within uncertainties across these energies. However, the CME signal disappears at both 7.7 and 200 GeV. These findings underscore that the observability of the CME critically depends on both the dynamic evolution of the magnetic field and the chemical freeze-out time of the partonic phase, which vary significantly with collision energy.
\end{abstract}

\maketitle

\section{Introduction}


The ground state of the QCD gauge field, known as the $\theta$-vacuum, is classified by an integer topological invariant — the Chern-Simons number ~\cite{Belavin:1975fg}. Transitions between distinct vacuum sectors, proceeding via instanton or sphaleron processes that tunnel through the intervening energy barriers, induce a violation of the axial current conservation. This is expressed through the anomaly relation
\begin{equation}
\partial^\mu j_\mu^5 = 2 \sum_f m_f \langle \bar \psi i \gamma_5 \psi \rangle_A - \frac{N_f g^2}{16\pi^2} F_{\mu\nu}^a \tilde F^{\mu\nu}_a,
\end{equation}
which results in a net chirality imbalance quantified by $N_5\equiv N_R-N_L=\int{d^4 x}\,\partial_{\mu}j^{\mu}_5$ ~\cite{Witten:1979vv,Veneziano:1979ec,Vicari:2008jw,Schafer:1996wv}. It has been suggested that local domains exhibiting $\mathcal{P}$- and $\mathcal{CP}$-odd symmetries could emerge in the environment of early universe or in the relativistic heavy-ion collisions.



Relativistic heavy-ion collision experiments conducted at Relativistic Heavy Ion Collider (RHIC) of BNL and  Large Hadron Collider (LHC) of CERN are capable of producing a deconfined state of matter known as the quark-gluon plasma (QGP). The environment of QGP in such collisions is realized to resemble that of the early universe. It is expected that the restoration of chiral symmetry accompanies deconfinement, rendering light quarks essentially massless and assigning them a well-defined chirality.  

In non-central heavy-ion collisions, the spectator nucleons at ultra-relativistic velocities generate an extremely strong magnetic field pointing out of the reaction plane, with its magnitude expected to fall within the range of $10^{17}$-$10^{19}$ Gauss ~\cite{Skokov:2009qp,Deng:2012pc}. Given a nonzero chiral chemical potential $\mu_5$, which characterizes a net chirality imbalance in the medium, a magnetic field $\mathbf{B}$ is predicted to give rise to a vector charge current $\mathbf{J}$ aligned with $\mathbf{B}$ inside the QGP. This phenomenon is known as the Chiral Magnetic Effect (CME) ~\cite{Fukushima:2008xe}, and is expressed as
\begin{equation}
{\mathbf {J}} = \frac{e^2}{2\pi^{2}}\mu_{5}{\mathbf {B}}.
\label{eq:CME_current}
\end{equation}
The CME-induced current causes positively and negatively charged particles to drift in opposite directions orthogonal to the reaction plane. As a result, the chiral anomaly and CME originating from the early stage of the collision translate into a measurable charge separation signal in the final hadronic freeze-out configuration ~\cite{Kharzeev:2015znc}.

The charge separation can be characterized by the azimuthal distribution $\phi$ of final-state charged particles relative to the reaction plane angle $\Psi_{\rm RP}$ ~\cite{Voloshin:2004vk},
\begin{equation}\label{eq:dndphi}
\frac{dN_{\pm}}{d\phi} \propto 1 + \sum_{n=1}^\infty 2v_{n}^{\pm}\cos n\Delta\phi + 2a_{1}^{\pm}\sin\Delta\phi,
\end{equation}
where $\Delta\phi \equiv \phi-\Psi_{\rm{RP}}$. The coefficients $v_n$ represent the momentum anisotropy which originates from the initial collision geometry and is transferred into the final state by hydrodynamic expansion ~\cite{Heinz:2013th}. The coefficients $a_{1}^{+}$ and $a_{1}^{-}$ are expected to have opposite signs as an observable of charge separation~\cite{Voloshin:2004vk}. Because QCD respects parity globally while permitting local violations, an experimental probe of the CME is designed to access charge-dependent fluctuations in $a_1$. Although $\langle a_1 \rangle$ vanishes when averaged over many events, the charge-splitting signal can be isolated through two-particle azimuthal correlations,
\begin{equation}\label{eq:gamma_delta}
\begin{split}
\gamma_{\alpha\beta} &\equiv \left\langle\cos(\phi_{\alpha} +\phi_{\beta} -2\Psi_{RP} )\right\rangle,  \\
\delta_{\alpha\beta} &\equiv \left\langle\cos(\phi_{\alpha} -\phi_{\beta})\right\rangle,
\end{split}
\end{equation}
where $\phi_{\alpha} ,\phi_{\beta} ,\Psi_{RP} $ denote the azimuthal angles of produced charged particles and the reaction plane, respectively, and $\alpha$ and $\beta$ represent either the positive or negative charges. Owing to the large background in these correlations, the differences between the opposite-sign (OS) and same-sign (SS) pairs are used to remove the common background in experiment, which are written as,
\begin{equation}
\begin{split}
    \Delta\gamma &=\gamma_{OS}-\gamma_{SS} = \kappa_{{\rm{bg}}} v_2 F - H, \\
\Delta\delta &=\delta_{OS}-\delta_{SS} = F + H,
\end{split}
\label{eq:FH}
\end{equation}
where $H$ and $F$ are the CME and background contributions, respectively, and $\kappa_{\rm{bg}}\approx 2-3$ is an adjustable parameter.
After canceling out $F$, we obtain
\begin{equation}
H = (\kappa_{\rm{bg}} v_2 \Delta\delta - \Delta\gamma)/(1+\kappa_{\rm{bg}} v_2).   
\label{eq:H_corr}
\end{equation}

The search for the chiral magnetic effect (CME) in high-energy heavy-ion collisions poses substantial challenges for signal identification~\cite{Kharzeev:2020jxw, Kharzeev:2024zzm}, as experimental measurements demand careful subtraction of large, intricate background contributions. Interpretations of the positive $\Delta\gamma$ measurements reported in Au+Au collisions signals observed in 200 GeV Au+Au collisions at RHIC~\cite{STAR:2021mii,STAR:2014uiw,STAR:2013ksd} and Pb+Pb collisions at LHC energies ~\cite{ALICE:2012nhw,ALICE:2012nhw} 
are complicated by flow-induced backgrounds, including effects from resonance decays~\cite{Voloshin:2004vk}, local charge conservation (LCC)~\cite{Schlichting:2010qia,Huang:2015oca}, and transverse momentum conservation (TMC)~\cite{Pratt:2010zn,Bzdak:2012ia}. Additionally, nonflow correlations~\cite{STAR:2000ekf,STAR:2019xzd} arising from particle clusters, resonances, jets, and dijets—sources uncorrelated with the reaction plane—can also generate positive   $\Delta\gamma$ contributions. Despite these difficulties, conclusive experimental searches for the quark-sector CME would illuminate key fundamental properties of QCD, particularly the yet-unobserved topological structures of the QCD vacuum, and offer a natural explanation for the cosmic baryon asymmetry~\cite{Kharzeev:2020jxw}.


Since the CME signal is predominantly generated during the earliest moments of a heavy-ion collision, subsequent final-state interactions tend to suppress its initial magnitude~\cite{Ma:2011uma}. One experimental strategy proposed to mitigate flow-associated backgrounds exploits subtle signal differences between two isobaric collision systems with nearly identical background conditions, namely $^{96}_{44}$Ru+$^{96}_{44}$Ru and $^{96}_{40}$Zr+$^{96}_{40}$Zr ~\cite{Deng:2016knn}. Nevertheless, the STAR isobar data at $\sqrt{s_{NN}} = 200$ GeV revealed no observable CME signature ~\cite{STAR:2021mii,STAR:2023gzg,STAR:2023ioo}, implying that background contributions dominate over any potential CME signal at this beam energy. Consequently, recent experimental efforts have shifted toward exploring a range of energies within the RHIC Beam Energy Scan Phase II (BES-II) program ~\cite{STAR:2025vhs,STAR:2025uxv}. The BES-II dataset features statistical precision improved by more than an order of magnitude relative to BES-I and leverages a dedicated detector for spectator-plane reconstruction, allowing for a more stringent probe of the CME signal.

The dominant background arises mainly from the emission pattern of final-state particles, governed by fluctuations in the initial collision overlap geometry (characterized by eccentricity) as well as the subsequent expansion of the QGP. Several approaches built upon event-shape engineering and event-shape selection have been established. These methods construct an event-shape observable using a kinematic region excluding particles of interest (POI), exploiting long-range flow correlations~\cite{Schukraft:2012ah,ALICE:2017sss,ALICE:2026cvp,CMS:2017lrw}. In this work, we adopt a recently proposed event-shape selection (ESS) technique~\cite{Xu:2023elq,Milton:2021wku}. This method simultaneously encodes the geometric and collective expansion features of each collision event and enables a more precise extrapolation toward the vanishing-
$v_2$ limit.

The AMPT model holds a distinctive advantage in simulating the partonic evolution and has been widely adopted for heavy-ion collision simulations. To describe QGP evolution in the presence of magnetic fields, we develop a partonic cascade framework dubbed chiral anomaly transport (CAT). The CAT framework extends the standard AMPT model by incorporating the fundamental chiral kinetic equation, which self-consistently conserves the charge current generated by the chiral anomaly and background magnetic field~\cite{Yuan:2023skl,Yuan:2024wpz}. The upgraded AMPT-CAT model successfully reproduces STAR measurements for 200 GeV Au+Au and isobaric collisions.

\section{Event Shape Selection}

The event shape selection method constitutes a novel strategy to search for the CME signal by suppressing the flow background entering Eq.\ref{eq:FH}. This method relies on three core steps. First, the chosen event-shape variable offers an event-by-event proxy for the initial collision geometry associated with particles of interest (POI). Second, for event classes sorted according to the event-shape variable, we evaluate the elliptic flow
$v_2$ and the CME observable $\Delta\gamma$ within each group, thereby establishing the
$\Delta\gamma$ versus $v_2$ dependence. Third, by extrapolating $\Delta\gamma$ to the zero-flow limit using the extracted correlation, flow-induced background contributions ideally vanish or are substantially suppressed, while the intrinsic CME signal can be reliably recovered.


Events can be categorized by the event-shape vector $\overrightarrow{q_2}$, defined as,
\begin{equation}
    \begin{split}
        q_{2,x}&=\frac{1}{\sqrt{N}}\sum_{i=1}^{N}\cos(2\phi_i)\\
        q_{2,y}&=\frac{1}{\sqrt{N}}\sum_{i=1}^{N}\sin(2\phi_i)
    \end{split}
\end{equation}
$\overrightarrow{q_2}$ provides a more robust and reliable extrapolation of $\Delta\gamma$ to the zero-flow limit compared to $v_2$ used in the event shape engineering method ~\cite{Schukraft:2012ah,Wen:2016zic}. By expanding the definition of $q_2$, we obtain ~\cite{Xu:2023elq} :
\begin{eqnarray}
q^2_2 & = & \frac{1}{N}
\biggl[\biggl(\sum^N_{i=1} \sin2\phi_i\biggr)^2 + \biggl(\sum^N_{i=1} \cos2\phi_i\biggr)^2 \biggr] \nonumber \\
& = & 1 + \frac{1}{N}\sum_{i\neq j} \cos[2(\phi_i - \phi_j)], \label{q2_definition}
\end{eqnarray}
from which the event-averaged value can be estimated as:
\begin{equation}
\langle q_2^2\rangle \approx 1 + N v_2^2\{2\}. \label{q2_average}
\end{equation}
Here, $v_2\{2\}$ denotes the elliptic flow extracted via two-particle correlations.
To properly account for finite multiplicity effects, the event shape variable in Eq.~(\ref{q2_definition}) is normalized as:
\begin{equation}
q^2_2 = \frac{\bigl(\sum^N_{i=1} \sin2\phi_i\bigr)^2 + \bigl(\sum^N_{i=1} \cos2\phi_i\bigr)^2 }{N(1+Nv_2^2\{2\})}.\label{eq:new_q}    
\end{equation}
While the particle multiplicity $N$ fluctuates on an event-by-event basis, $v_2\{2\}$ remains a static, ensemble-averaged constant. This normalization correction becomes particularly prominent in high-multiplicity scenarios.
 
Conventionally, $q_2$ is constructed using single particles of interest (POIs) under the assumption that single-particle $q_2$ and $v_2$ primarily reflect the initial spatial geometry of the collision system. However, since the correlator $\gamma$ inherently correlates the reaction plane with particle pairs rather than isolated single particles, the anisotropy of particle pairs may be more directly linked to background contributions. For example, for backgrounds originating from flowing resonances, the collective flow $v_2$ and shape variable $q_2$ of the parent resonances are physically more relevant than those of their respective decay daughters. 

To address this issue, we define the pair azimuthal angle $\phi^{\rm p}$, which is constructed from the total momentum vector of each pair of POI pair, regardless of the existence of a physical parent resonance. The corresponding pair elliptic flow $v_{2,\rm pair}$ and pair event-shape variable $q^2_{2,{\rm pair}}$ are given by:
\begin{eqnarray}
v_{2,\rm{pair}} &=& \langle \cos(2\phi^{\rm p} - 2\Psi_{\rm {RP}}) \rangle, \label{eq:pairv2}\\
q^2_{2,{\rm pair}} &=& \frac{\bigl(\sum^{N_{\rm pair}}_{i=1} \sin2\phi_i^{\rm p}\bigr)^2 + \bigl(\sum^{N_{\rm pair}}_{i=1} \cos2\phi_i^{\rm p}\bigr)^2 }{N_{\rm pair}(1+N_{\rm pair}v_{2,{\rm pair}}^2\{2\})}, \label{eq:pairq2}
\end{eqnarray}
respectively. Since the "unmixed" frameworks--which combine either single $q_2^2$ with single $v_2$, or pair $q_2^2$ with pair $v_2$-- extract both observables from identical sets of particles or particle pairs, they are subject to inherent mathematical correlations illustrated by Eqs.~(\ref{q2_definition}) and (\ref{q2_average}). As a result, extrapolations toward the zero-flow limit within such frameworks may suffer from systematic biases, which yield a residual background contributions. In contrast, "mixed" strategies, where single $q_2^2$ is paired with pair $v_2$ (or vice versa), impose stronger statistical independence between the event-shape and elliptic-flow observables, potentially enabling a cleaner and more reliable extrapolation. Existing STAR analysis ~\cite{STAR:2025vhs,STAR:2025uxv} and theoretical studies ~\cite{Xu:2023elq} indicate that the configuration $v_{2,single}\{ q^2_{2,pair}\}$ exhibits better performance compared with other configurations.

\section{ The AMPT-CAT Model Setup}

\subsection{AMPT model}
In this paper, we employ the AMPT model incorporating the chiral anomaly transport (CAT) module to study the CME via the ESS method for Au+Au collisions at BES-II energies. 

The AMPT model consists of four modules to simulate the four stages of heavy ion collisions: initial state, parton cascade, hadronization, and hadron rescatterings. There exist two variants of the AMPT model: the default version and the version with a string-melting version. In both versions of AMPT, the initial state of collision generated by the HIJING contains particles originating from hard and a soft processes. The hard and soft processes are classified according to whether the momentum transfer exceeds a cutoff momentum $p_0$, producing strings or minijet, respectively.  In the default AMPT version, partons are produced by the minijet following the PYTHIA program, while excited strings fragment into hadrons via the Lund JETSET model. In the string-melting version, strings and minijets melt into partons, yielding a higher parton density compared with the default setup. The parton cascade evolution can be modeled by the original ZPC module in standard AMPT or, as adopted in the present work, the chiral anomaly transport (CAT) \cite{Yuan:2023skl,Yuan:2024wpz}  module. The chiral anomaly transport module is developed to describe QGP evolution under magnetic fields incorporating chiral anomaly effects, serving as a replacement for the original ZPC model, which only accounts for elastic binary parton collisions. Hadronization is realized via a simple quark combination model, and subsequent hadron rescatterings are described by a hadron transport model.

In this simulation, the initial parameters include the Lund string
fragmentation parameters $a=0.55, b=0.15\ \text{GeV}^{-2}$,strong coupling constant $\alpha_s=0.33$ and parton screening mass $m_s=2.265\ \text{fm}^{-1}$ at $\sqrt{s_{NN}}=$200, 27 and 19.6 GeV, $m_s=2.3\ \text{fm}^{-1}$ at $\sqrt{s_{NN}}=$11.5 GeV and $m_s=2.5\ \text{fm}^{-1}$ at  $\sqrt{s_{NN}}=$ 9.2 and 7.7 GeV which can reasonably reproduce many experimental observables such as rapidity distributions, $p_T$ spectra, and anisotropic flows.

 
    

    

    

\subsection{Nuclear Thickness}
The finite nuclear thickness at lower energies are considered in the AMPT model~\cite{Wang:2021owa}. The starting time $t=0$ is the moment when the edge of projectile and target nuclei contact each other and the collision time $t_c$ is defined as the moment when nucleon collides to another nucleon . For a nucleon at position $(x_i,y_i,z_i)$, the thickness length $l$ of colliding nuclei is, 
\begin{equation}
    l(x_i,y_i,b)=2\sqrt{R^2-(x_i \pm b/2)^2-y_i^2}
\end{equation}
where $b$ is the impact parameter, $R$ is the hard-sphere radius of colliding nuclei, and $\pm$ applies to projectile or target nucleons, respectively. The time $t_e$  is the moment when nucleon enters the colliding nuclei in the center-of-mass frame and can be calculated as follow,

\begin{equation}
   t_e(x_i,y_i,z_i,b)=\frac{\sqrt{R^2-b^2/4}-[l(x_i,y_i,b)/2+z_i]}{2\text{sinh}y_{CM}}
\end{equation}

The projectile nucleon takes a duration of $d_t = l / (2\sinh y_{\text{CM}})$ to completely traverse the colliding nuclei, exiting at $t_e(x_i, y_i, z_i, b) + d_t$. The nucleon interaction time, which corresponds to the production time of parent hadron $t_{H} \in [t_e,t_e+dt] $ is is sampled following~\cite{Wang:2021owa,Lin:2017lcj}. The longitudinal coordinate of the parent projectile or target nucleon at the interaction time is given by: 
\begin{equation}
   z_H=z_i+t_H \text{sinh}y_{CM}
\end{equation}
After parent hadron production, partons are generated by string melting after a formation time $t_f=E_H/m_{T,H}^2$where $E_H$ and $m_{T,H}$ denote the energy and transverse mass of the parent hadron, respectively. Finally, the partons originating from the same parent hadron are produced at the same space-time coordinates, $t_q = t_H + t_f$ and $z_q = z_H$.

\subsection{Magnetic Field}


 To simulation the evolution of magnetic field, we use an algebraic expression for the tangential component $B_{\phi}$ considering conductivity $\sigma$ ~\cite{Li:2016tel, Gursoy:2014aka},

\begin{eqnarray}
B_{\phi}\left(t,\mathbf{x}\right) & = & \alpha_{EM}\sum_i \frac{v_i\gamma x_{T,i}}{\Delta_i^{3/2}}\left(1+\frac{\sigma v\gamma}{2}\sqrt{\Delta_i}\right)e^{A_i},\label{eq:mag-phi}
\end{eqnarray}
where $i$ denotes i-st spectator proton and we define $\Delta_i\equiv\gamma^{2}(v_it-z_i)^{2}+x_{T}^{2}$, $A_i\equiv(\sigma v\gamma/2)[\gamma(vt-z)-\sqrt{\Delta}]$ and $x_{T,i}^2=x_i^2+y_i^2$. Considering the absence of QGP before the collision and that the conductivity of vacuum should be zero, we describe the time dependence of conductivity as a $\theta$ function~\cite{Li:2023tsf},


\begin{eqnarray}
    \sigma(t)=\sigma_0\theta(t-t_e)
\end{eqnarray}
where $t_e$ is the moment when spectator enters the colliding nuclei and we adopt a conductivity parameter $\sigma_0 = 5.8\ \text{MeV}$ as the conductivity of the QGP, following Ref.~\cite{Tuchin:2013ie}.

\begin{figure}[tb]
\centering
\includegraphics[width=0.43 \textwidth]{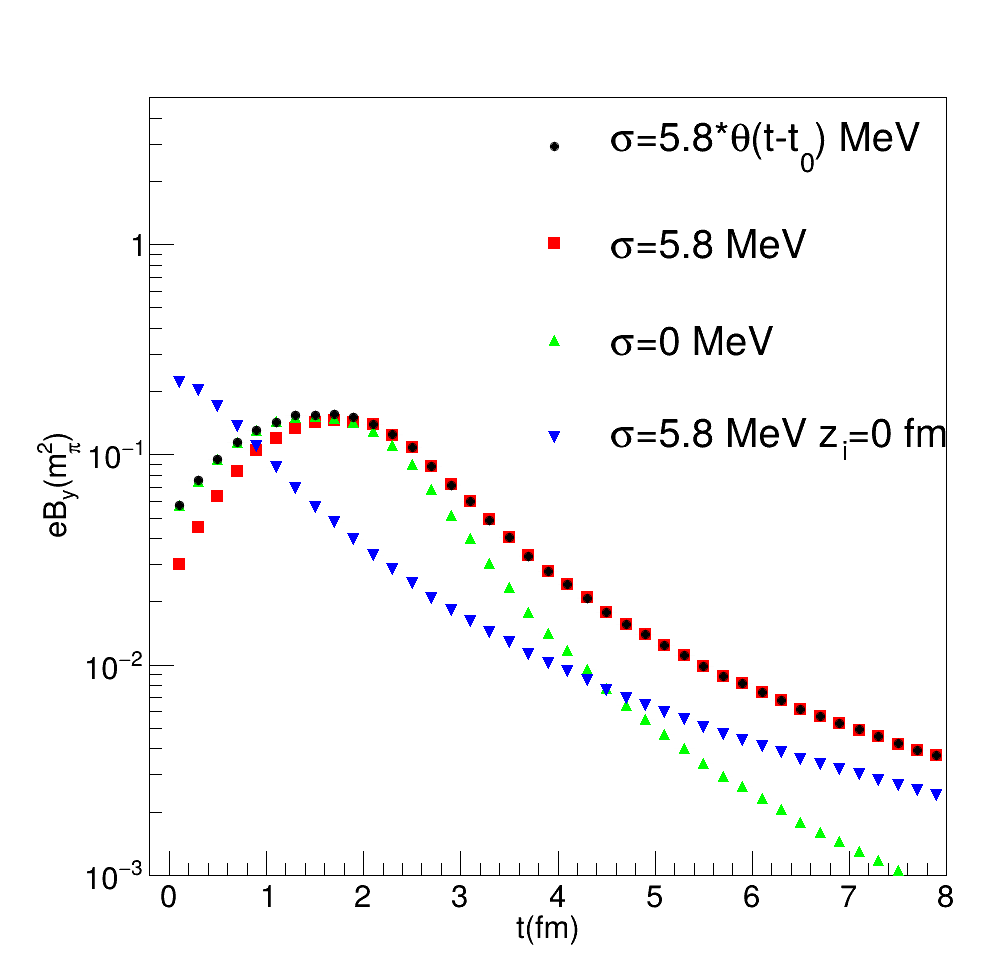}
\caption{Time evolution of magnetic field $eB_{y}(t)$ in Au + Au collisions at $r=(0,0,0)$, $b=8 fm$ and $\sqrt{s_{NN}}=11.5\ \text{GeV}$. }
\label{fig:eb-t-11.5}
\end{figure}

\begin{figure}[tb]
\centering
\includegraphics[width=0.43 \textwidth]{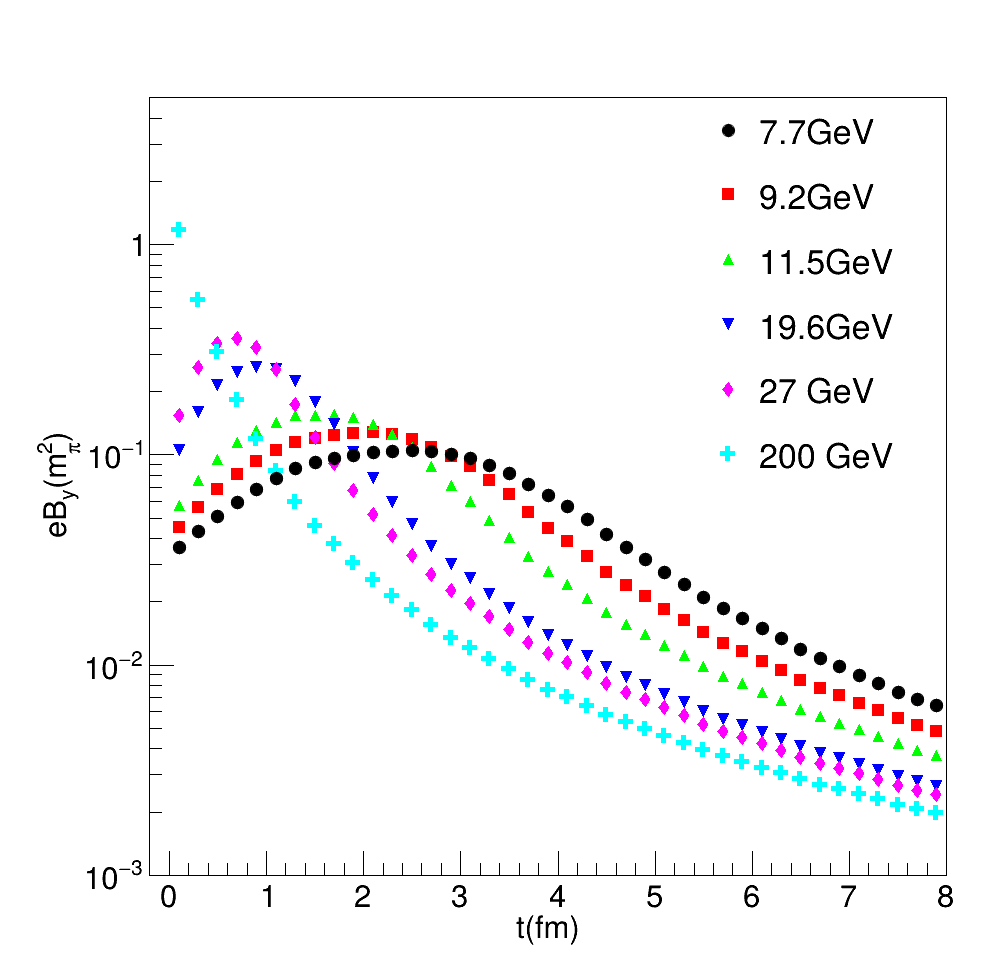}
\caption{Time evolution of magnetic field $eB_{y}(t)$  in Au + Au collisions at $r=(0,0,0)$, $b=8 fm$ and $\sqrt{s_{NN}}=7.7-200\ \text{GeV}$. }
\label{fig:eb-t}
\end{figure}

Fig.~\ref{fig:eb-t-11.5} presents the time evolution of $eB_y(t)$ at the center of the collision for $\sqrt{s_{NN}} = 11.5\ \text{GeV}$, under four different conditions: $\sigma = 5.8 \cdot \theta(t-t_e)\ \text{MeV}$ (black dot), $\sigma = 5.8\ \text{MeV}$ (red dot), $\sigma = 0\ \text{MeV}$ (green dot), and $\sigma = 5.8\ \text{MeV}$ when ignore nuclear thickness and set $z_i=0$ for all spectators (blue dot). Compared to the $\sigma = 0\ \text{MeV}$ case, the constant conductivity $\sigma = 5.8\ \text{MeV}$ yields a smaller initial $eB_y(t)$ but a longer decay time. In the $\sigma = 5.8 \cdot \theta(t-t_e)\ \text{MeV}$ case, $eB_y(t)$ initially behaves similarly to the $\sigma = 0\ \text{MeV}$ case, and approaches that of the $\sigma = 5.8\ \text{MeV}$ case at large times.In the first three cases, where the finite thickness of the nuclei is taken into account, the evolution curves of the magnetic field exhibit an initial rise followed by a decay. In contrast, when aignore nuclear thickness (blue curve), the magnetic field reaches its maximum at the initial time.

Fig.~\ref{fig:eb-t} illustrates the time evolution of the out-of-plane magnetic field $eB_y(t)$ at the center of the collision ($r = (0,0,0)$) for an impact parameter $b = 8$~fm, considering collision energies $\sqrt{s_{NN}} = 7.7-200\ \text{GeV}$ in this simulation. The maximum magnitude of $eB_y(t)$ increases with collision energy. However, the decay rate becomes slower at lower collision energies. The prolonged lifetime of the magnetic field at lower energies arises from the fact that the field is predominantly generated by the thickness of nuclei and the spectator moving with velocities proportional to the collision energy. Moreover, the peak position shifts to later times at lower energies due to the reduced velocity of the spectators and the finite thickness of the nuclei. At lower collision energies, the spectators take longer to pass the center after the collision begins. But at 200 GeV, the Lorentz-contracted thickness of the nucleus is near zero, the peak is at the  beginning.

\subsection{Elliptic Flow}

\begin{figure}[tb]
\centering
\includegraphics[width=0.45 \textwidth]{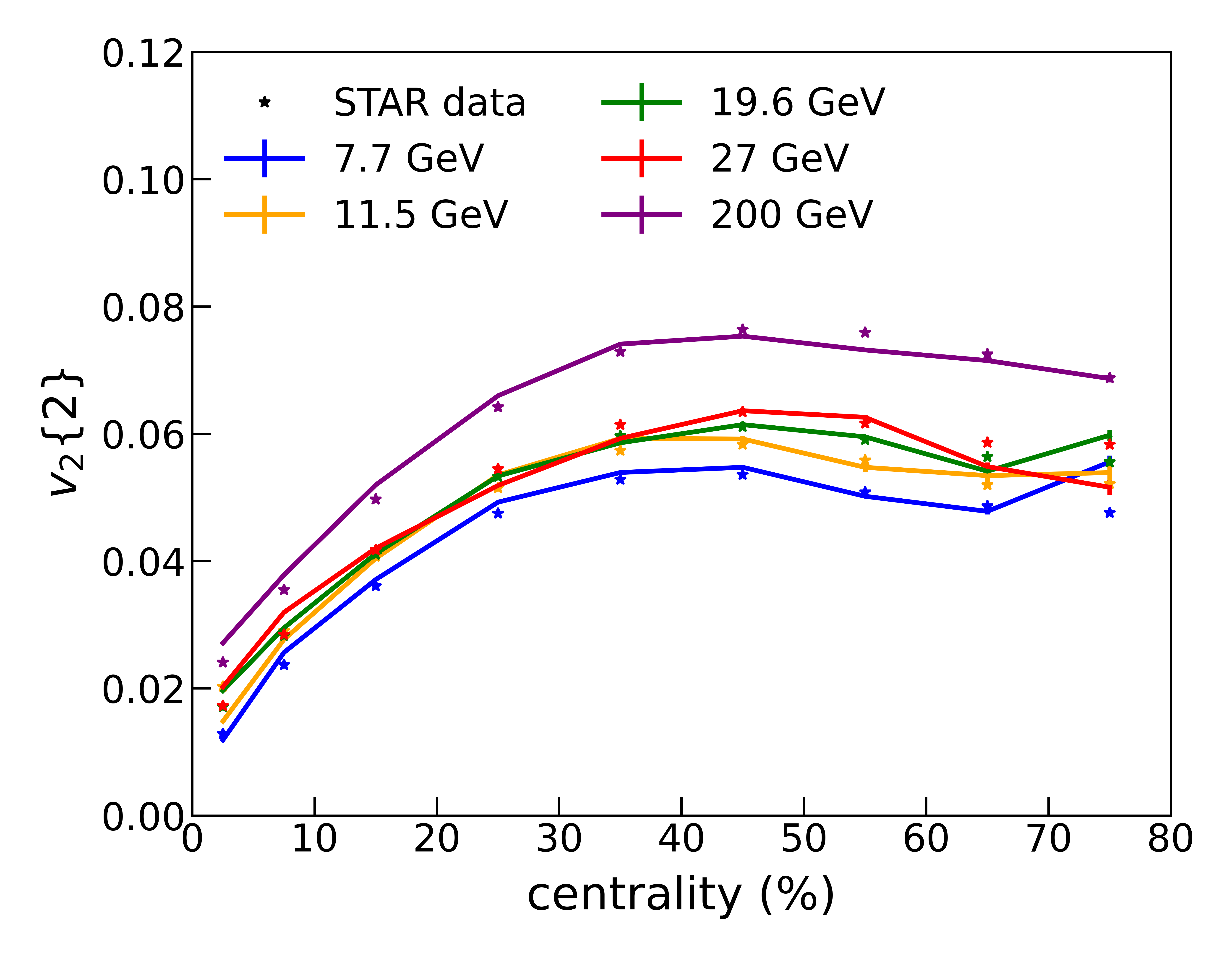}
\caption{The centrality dependence of elliptic flow $v_2\{2\}$ from AMPT (solid lines) compared with STAR data (colorful circles) ~\cite{STAR:2012och} in Au + Au collisions at $\sqrt{s_{NN}}=7.7,19.6,27,200\ \text{GeV}$. The POI are charged hadrons within $0.2<p_{T}<1.4 \ \text{GeV/c}$ and $|\eta|<1$.}
\label{fig:v22}
\end{figure}

The elliptic flow term is a great background contribution in the CME observation $\Delta\gamma$. In the simulation, it is necessary to make sure the elliptic flow of model is consistent to the experiment data. We take the advantage of the two-particle correlation $v_2\{2\}$ which is  independent to the reaction plane, to reduce variations from different reaction plane determinations and low reaction plane resolutions. Figure \ref{fig:v22} presents the centrality dependence of $v_2\{2\}$ for charged hadrons within $|\eta|<1,p_T>0.2\ \text{GeV/}c^2$ in Au+Au collisions at 200 GeV and BES-II energies, comparing AMPT model results with STAR experimental data ~\cite{STAR:2012och}. The AMPT model provides a good description of the STAR data over these energies in the most central and mid-central collisions, suggesting that the obtained flow background is credible. 

 In the present work,to maximize event plane resolution and minimize non-flow effect  ~\cite{STAR:2005btp}, we take the first-order spectator plane as the reaction plane $\Psi_{RP}=\Psi_{SP}$ in the model as same as STAR ESS analysis~\cite{STAR:2025vhs,STAR:2025uxv} , which is defined as,
\begin{equation}
    \Psi_{\text{SP}}=\arctan(\frac{\left \langle r_s \sin(\phi) \right \rangle }{\left \langle r_s \cos(\phi) \right \rangle})
\end{equation}
where $r_S$ denotes the displacement of spectator protons relative to the field point $r=(0,0,0)$ and $\phi$  represents the azimuthal angle in the transverse plane. In the STAR ESS analysis, only charged mesonss with in $p_T>0.2\text{GeV/}c^2, p<1.4\ \text{GeV/}c^2$ ($p<1.3\ \text{GeV/}c^2$ at 7.7 GeV) are considered and the elliptic flow is smaller than $v_2\{2\}$ which consider all charged hadrons~\cite{STAR:2012och}. 


\subsection{Chiral Anomaly}


As shown in Fig.~\ref{fig:eb-t}, the maximum magnitude of the magnetic field is $ O(m_{\pi})\sim 10^{-1} \text{ GeV} $. Consequently, if the magnitude of the chiral chemical potential $ \mu_5 $ exceeds $ 10^{-1} \text{ GeV} $, the chiral charge density of the quark-gluon plasma can be evaluated under the weak magnetic field limit~\cite{Fukushima:2008xe}:
\begin{equation}
    n_5 = \frac{\mu_5^3}{3\pi^2} + \frac{\mu_5}{3}\left(T^2 + \frac{\mu^2}{\pi^2}\right),
\end{equation}
where $ \mu_5 $ denotes the quark chemical potential. In the simulation, the entire spatial volume is divided into many cells of size $ 1 \times 1 \times 0.5 \text{ fm}^3 $. For each cell $ i $, the local chiral chemical potential $ \mu_{5,i} $ and chiral charge density $ n_{5,i} $ are computed from the local magnetic field $ eB_i $, the local temperature $ T_i $, and the global chemical potential $ \mu $. The total global chiral quark imbalance is then obtained by summing over all cells:
\begin{equation}
    N_5 = \int d^3x \, n_5 = V_{\text{cell}} \sum_i n_{5,i},
\end{equation}
where $ V_{\text{cell}} $ is the volume of a single cell.

\begin{figure}[htbp]
    \centering
    \hspace*{-.5cm}\includegraphics[width=0.9\linewidth]{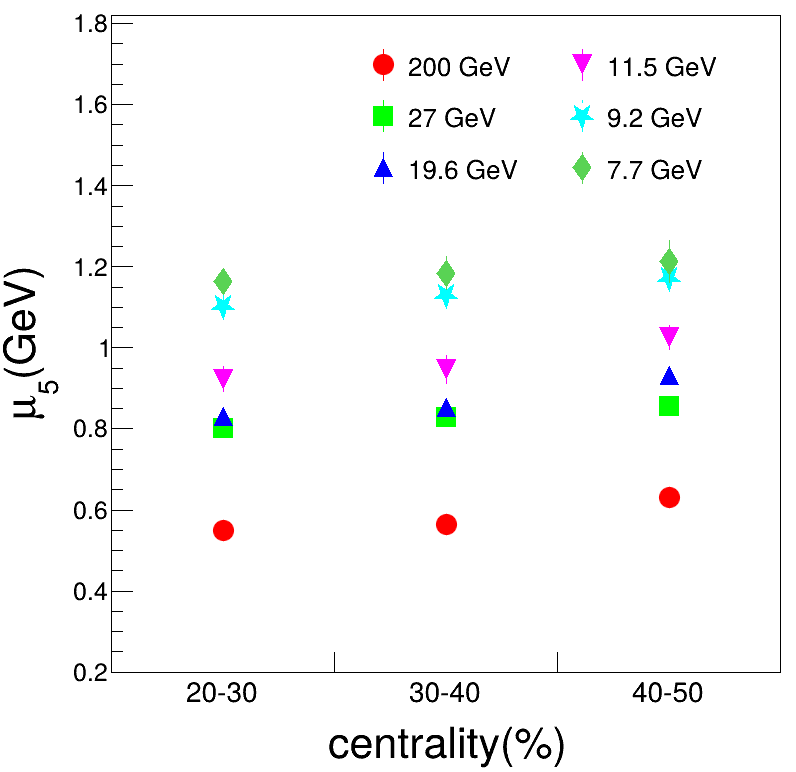}
    \caption{The centrality dependence of chiral chemical potential $\mu_5$ at $7.7-200$ GeV.}
    \label{fig:input_mu5}
\end{figure}

Assuming that the magnitude of $\mu_5$ depends on the magnetic field, temperature, and baryon chemical potential, its absolute value is parametrized as:
\begin{equation}
    |\mu_5| = a\sqrt{|eB|} + bT + c\mu_B,
    \label{eq:mu5_definition}
\end{equation}
while its sign is randomly assigned to be positive or negative on an event-by-event basis. Since the CME signal is proportional to the magnitude of the chiral chemical potential $|\mu_5|$, a parameter scan is performed using AMPT with the basic AMPT parameters already constrained by the experimental data on elliptic flow and meson yields. By constraining these parameters against the STAR $\Delta\gamma$ data, we obtain $a = 5.51$, $b = 0.67$, and $c = 1.4$. Figure~\ref{fig:input_mu5} presents the resulting $|\mu_5|$ values used in this study across different centrality classes for $\sqrt{s_{\mathrm{NN}}} = 7.7\text{--}200~\text{GeV}$.


\section{Results}
As shown in Eq.~\ref{eq:CME_current}, the charge current of CME is directly proportional to the chiral chemical potential $\mu_5$. As consequence,when $\mu_5=0$, the CME current vanishes and the observable $\langle N_{\text{part}}\Delta\gamma \rangle$ receives contributions solely from background sources, such as flow-driven correlations and local charge conservation effects. 
Since the chiral chemical potential does not modify the collective flow or the bulk multiplicity in a significant way, the background contribution can be considered unchanged between the finite and zero $\mu_5$ cases. For non-zero $\mu_5$, the same background contributions remain present, but an additional CME signal arises from the chiral anomaly. 
Under this assumption, the pure CME signal can be cleanly extracted by subtracting the $\mu_5=0$ baseline from the finite-$\mu_5$ result, which named "direct" method,
\begin{equation}
    \langle N_{\text{part}}\Delta\gamma_{CME}\rangle\{direct\}=\langle N_{\text{part}}\Delta\gamma_{|\mu_5|>0}\rangle-\langle N_{\text{part}}\Delta\gamma_{\mu_5=0}\rangle
\end{equation}

\begin{figure}[tb]

    \includegraphics[width=0.49\textwidth]{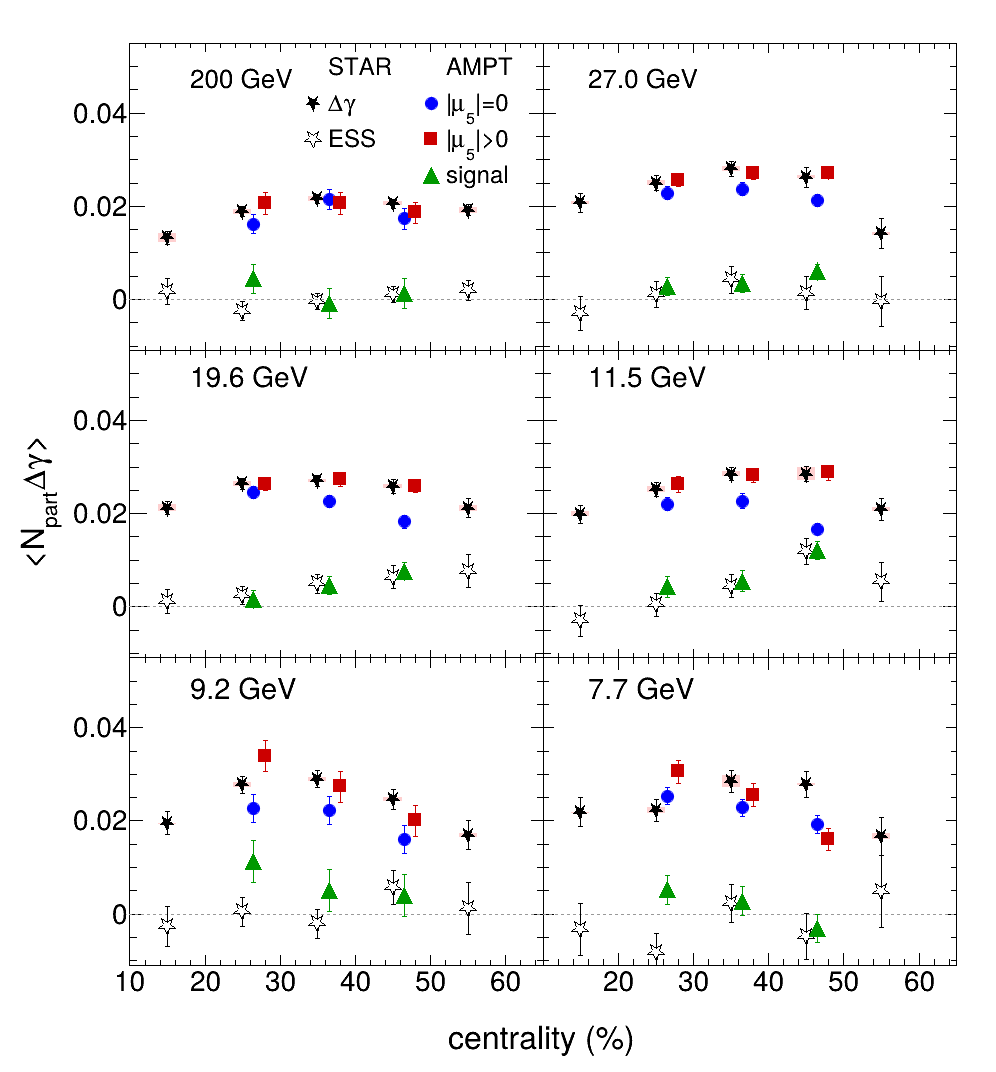}
\caption{The centrality dependence of $N_{\mathrm{part}}\Delta\gamma$ with non-zero $\mu_5$ (blue lines) and zero $\mu_5$ (orange lines) and the CME signal (green lines) extracted by AMPT compare with STAR data (black circles) in Au + Au collisions at various beam energies. The POI are charged mesons within $0.2<p_{\mathrm{T}}<1.4$ GeV,$|\eta|<0.2$ using the same kinematic selection as in the STAR analysis.}
\label{fig:Npart_gamma_combined}
\end{figure}

This analysis focuses on mid-central collisions in 7.7-200~GeV, where the magnetic field is expected to be strong and induce a significant CME signal. 
Fig.~\ref{fig:Npart_gamma_combined} presents the centrality dependence of $\langle N_{\text{part}}\Delta\gamma \rangle$ and the extracted CME signal $\langle N_{\text{part}}\Delta\gamma_{\mathrm{CME}} \rangle$ from both the AMPT simulations (with and without chiral anomaly) and the STAR experimental data at various collision energies.
The AMPT calculations, which incorporate the chiral anomaly, can reasonably describe the STAR data points, and the magnitude of the extracted CME signal is qualitatively consistent with the ESS results reported by STAR. 
At $\sqrt{s_{NN}} = 7.7~\text{GeV}$, the AMPT $\langle N_{\text{part}}\Delta\gamma \rangle$ in the $40\text{--}50\%$ centrality range is smaller than the STAR data, indicating that AMPT slightly underestimates the background contribution to this observable.
At 9.2~GeV,$\langle N_{\text{part}}\Delta\gamma \rangle$ of AMPT is consistent with STAR data, but the AMPT CME signal is slightly larger than the STAR ESS measurement. 
These differences could stem from the imperfect description of strange meson yields, $p_T$ spectra in AMPT at these low energies and absent background contributions, such as local charge conservation (LCC), that are not fully considered in the AMPT model. 
At 11.5 and 19.6~GeV, both the AMPT simulation and the STAR ESS analysis exhibit distinct separations between the total and the background-only $\langle N_{\text{part}}\Delta\gamma \rangle$, with a larger difference compared to that at lower energies. 
At 27~GeV, AMPT reproduces the STAR $\langle N_{\text{part}}\Delta\gamma \rangle$ data well. 
In the AMPT results, the CME signal shows the same tendency as at 11.5 and 19.6~GeV and is consistent with the STAR ESS results in the $20-40\%$ centrality class; however, the STAR ESS result in the $40-50\%$ centrality class decreases to near zero.
At the highest collision energy, 200~GeV, the CME signal becomes much smaller than that observed in the $11.5-27$~GeV range.

\begin{figure}[!htb]
    \centering
    \includegraphics[width=1\linewidth]{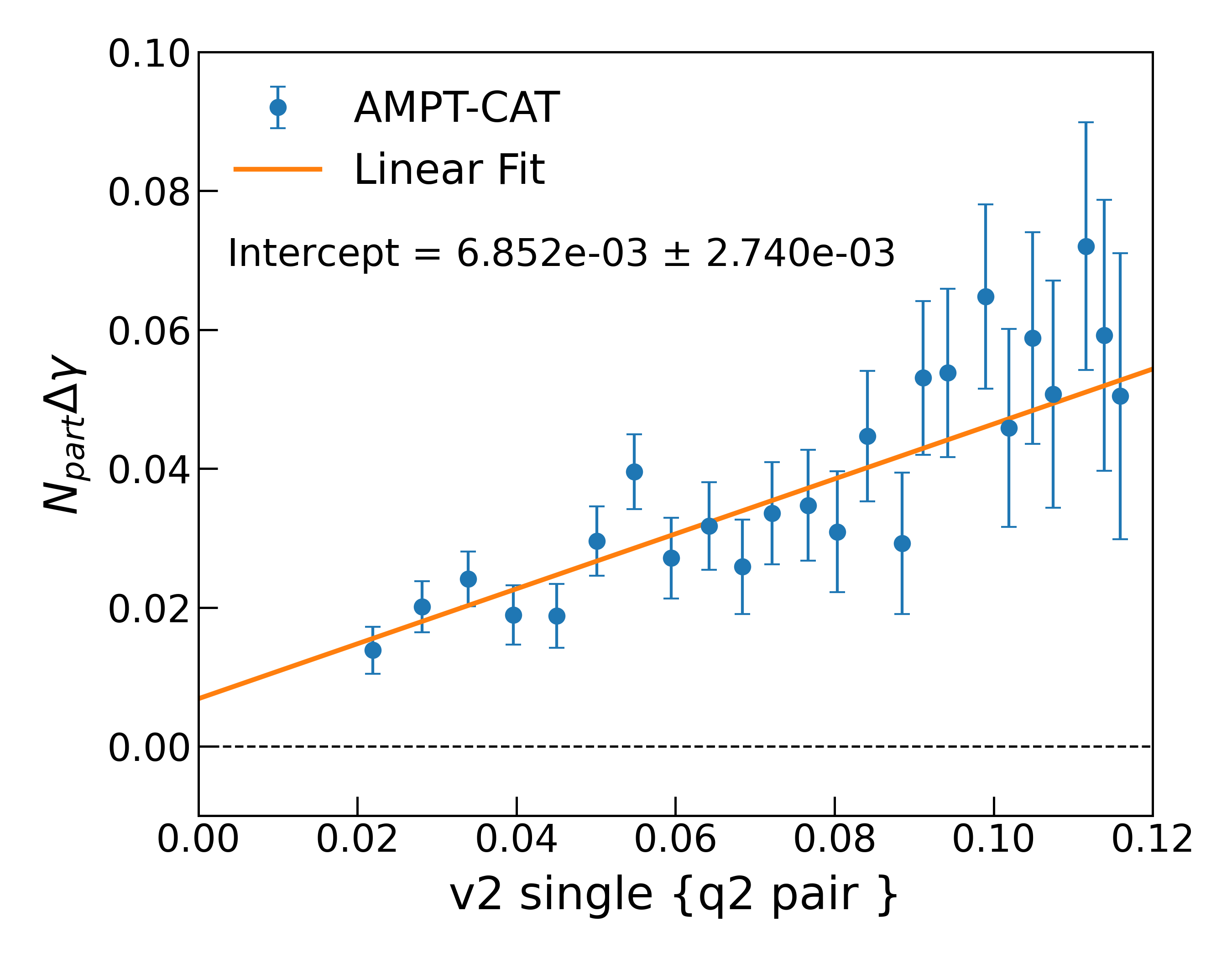}
    \caption{The ESS approaches that extrapolate $N_{part}\Delta\gamma$ to zero $v_2\{POI\}$ of signal interesting particle with events classified $q_2\{PPOI\}$ of pair of interesting particle in 30-40$\%$ Au+Au collision at 19.6 GeV with $|mu_5|>0$. The linear fits (orange lines) are used to extract the y-intercepts.}
    \label{fig:ese_19.6_30-40}
\end{figure}


In the ESS approach applied in the AMPT simulation, $v_{2}\{\mathrm{POI}\}$ denotes the elliptic flow of particles of interest, and $q_{2}\{\mathrm{PPOI}\}$ is constructed from particle pairs, following the same procedure as in the STAR analysis. 
An example of the ESS analysis in $30-40\%$ central Au+Au collisions with the CME at 19.6~GeV is shown in Fig.~\ref{fig:ese_19.6_30-40}, illustrating a linear dependence of $\langle N_{\text{part}}\Delta\gamma \rangle$ on $v_{2}$. 
The intercept of the linear fit to $\langle N_{\text{part}}\Delta\gamma \rangle$ as a function of $v_{2}$ yields the CME signal extracted by the ESS method. 
The ESS CME signal from AMPT in this centrality bin is $(4.273 \pm 2.582) \times 10^{-5}$, which is consistent with the STAR result of $(4.9 \pm 2.0) \times 10^{-5}$.


\begin{figure}[tb]
\centering
    {\includegraphics[width=0.49\textwidth]{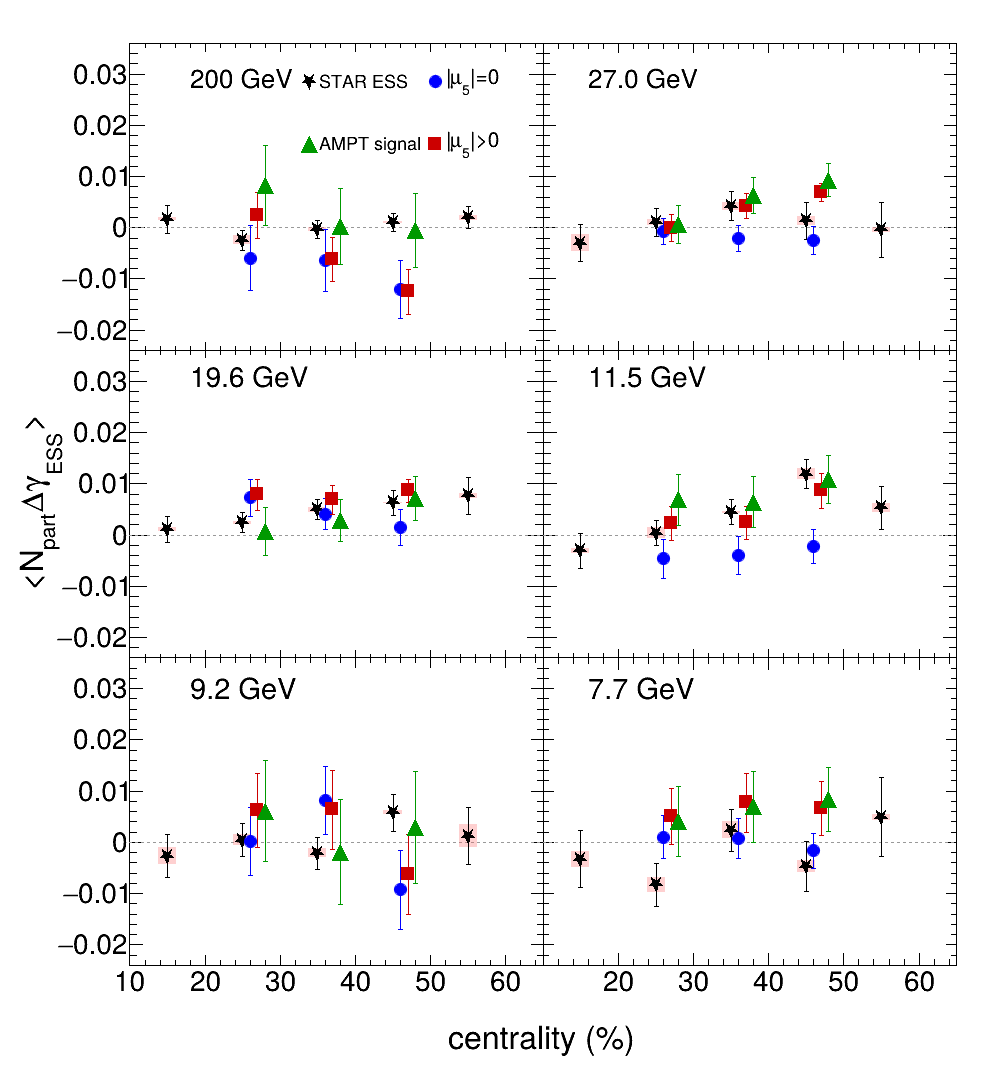}}

\caption{The centrality dependence of the intercept of $N_{part}\Delta\gamma$ using ESS approach with non-zero $\mu_5$ (blue lines) and zero $\mu_5$ (orange lines) compared with STAR ESS results (black circles) in Au + Au collisions at various beam energies.}
\label{fig:ESS_Npart_gamma_combined}
\end{figure}


Fig.~\ref{fig:ESS_Npart_gamma_combined} shows the centrality dependence of $\langle N_{\text{part}}\Delta\gamma_{ESS} \rangle$ with or without the chiral anomaly in the AMPT simulation, compared with the STAR data. 
To validate the ESS method, we test whether ESS yields a vanishing signal in a controlled, CME-free sample. However, as shown in Fig.~\ref{fig:ESS_Npart_gamma_combined}, the ESS results from AMPT at $\mu_5 = 0$ do not identically vanish, indicating the presence of residual background contributions that are not fully eliminated by the procedure. Consequently, in the ESS approach, the CME signal is extracted from the difference between $\langle N_{\text{part}}\Delta\gamma_{\mathrm{ESS}} \rangle$ obtained with non-zero and zero chiral chemical potential.
At 7.7~GeV, the ESS results with $\mu_5 = 0$ are consistent with zero, while the results with non-zero $\mu_5$ are positive, even though the STAR data points are negative in this centrality range. The large differences suggest that the background effects in the ESS method may differ significantly between the experimental data and the AMPT simulation, potentially due to the imperfect description of particle spectra and non-flow correlations in AMPT at 7.7~GeV.
At 200~GeV, the ESS result in the $20-30\%$ centrality range is larger than the corresponding STAR measurement, indicating a possible overestimation of the CME signal by AMPT or an incomplete cancellation of background contributions at high collision energies using ESS method.



\begin{figure}[tb]
\centering
\includegraphics[width=0.45 \textwidth]{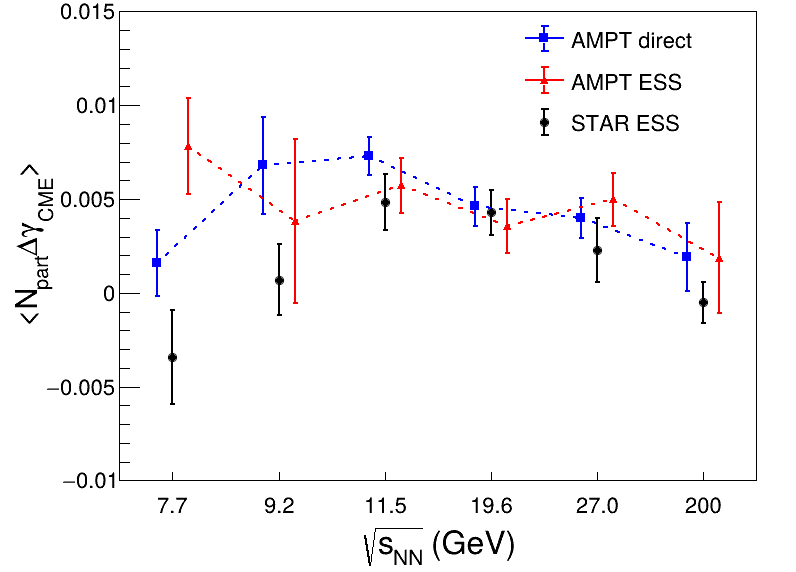}
\caption{The energy dependence of $N_{part}\Delta\gamma_{CME}$ extracted fromAMPT simulation in $20-50\%$ centrality comapred with STAR ESS results, where $N_{part}\Delta\gamma_{CME}\{direct\}=$ $N_{part}\Delta\gamma_{|\mu_5|>0}-N_{part}\Delta\gamma_{\mu5=0}$ and $N_{part}\Delta\gamma_{CME}\{ESS\}$ are the intercepts in ESS approach.}

\label{fig:Ng_ene_ess}
\end{figure}
\begin{figure}[tb]
\centering
\includegraphics[width=0.45 \textwidth]{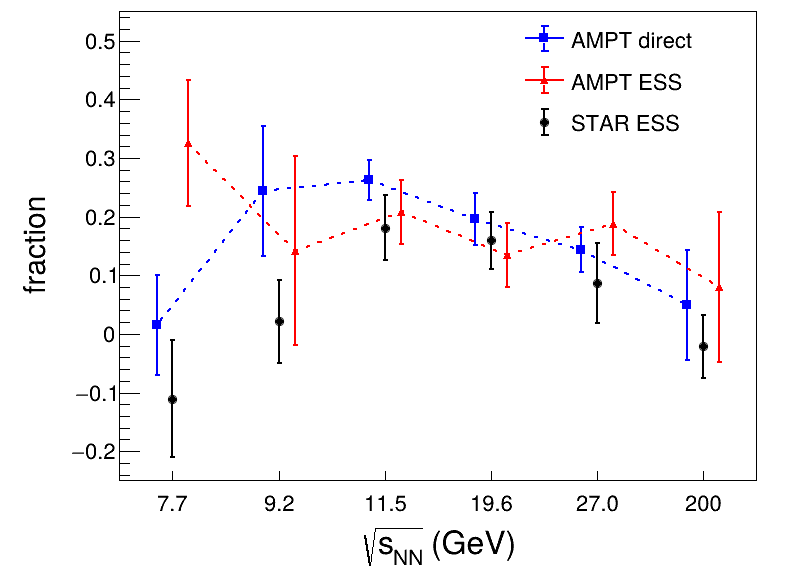}
\caption{The energy dependence of CME fraction $f=\Delta\gamma_{CME}/\Delta\gamma$ in $20-50\%$ centrality extracted from AMPT directly and by ESS method simulation compared with STAR data.}
\label{fig:Ng_ene_compare}
\end{figure}

Fig.~\ref{fig:Ng_ene_ess} shows the collision energy dependence of the directly and ESS extracted CME signal in $20-50\%$ central Au+Au collisions. And Fig.~\ref{fig:Ng_ene_compare} shows the collision energy dependence of the CME fraction $f = \Delta\gamma_{\mathrm{CME}} / \Delta\gamma_{\mathrm{tot}}$ obtained from the direct subtraction method and the ESS method. 
The CME fractions extracted by the two methods are consistent with each other within uncertainties at 9.2, 11.5, 19.6, 27, and 200~GeV. 
At 7.7~GeV, the observable $\langle N_{\text{part}}\Delta\gamma \rangle$ suggests CME fraction is much small, but the ESS method yields a significantly larger CME fraction. 
This discrepancy may arise from non-flow correlations that are not fully accounted for by the ESS procedure.
The chiral anomaly produces a significant CME signal in the energy range $\sqrt{s_{\mathrm{NN}}} = 9.2-27$~GeV, while the CME signal is consistent with zero at 7.7 and 200~GeV. 
At 200~GeV, the CME signal is small despite the magnetic field being the strongest among all collision energies considered. This can be understood by small chiral chemical potential, weak strength of magnetic field in the QGP phase~\cite{Yan:2021zjc} and depression of strong hadronic interaction.  
At 7.7~GeV, the CME signal also vanishes. 
In this case, the energy density of the produced QGP is relatively low, and the partonic phase has a shorter lifetime. Quarks hadronize and freeze out quickly after the collision, providing a very narrow time window for the CME current to develop. 
As a result, the observable $\langle N_{\text{part}}\Delta\gamma \rangle$ becomes insensitive to the chiral anomaly, even if a great $\mu_5$ is introduced in the initial state. 
These two scenarios highlight that the observability of the CME depends not only on the presence of a chiral imbalance but also critically on the lifetime of the magnetic field and the partonic phase, both of which vary strongly with collision energy.

The background coefficient $\kappa_{bg}$ quantifies the coupling between elliptic flow and the CME observable $\Delta\gamma$ and used to be set as 2.5, defined as,
\begin{equation}
    \kappa_{bg}=\frac{\Delta\gamma+H}{v_2(\Delta\delta-H)}
    \label{eq:k_def}
\end{equation}
where $H$ is the CME contribution term. If we consider the simulation with zero chiral anomaly can describe all the background effect, Eq.\ref{eq:k_def} can be written as,
\begin{equation}
    \kappa_{bg}\{direct\}=\frac{\Delta\gamma_{bkg}}{v_2\Delta\delta_{bkg}}=\frac{\Delta\gamma_{|\mu_5|=0}}{v_2\Delta\delta_{|\mu_5|=0}}
        \label{eq:k_class}
\end{equation}

Assuming that the ESS method eliminates all flow background and that $H = -\Delta\gamma_{\mathrm{ESS}}$, Eq.~\ref{eq:k_def} can be expressed as:
\begin{equation}
    \kappa_{bg}\{ESS\}=\frac{\Delta\gamma_{tot}-\Delta\gamma_{ESS}}{v_2(\Delta\delta_{tot}+\Delta\gamma_{ESS})}
        \label{eq:k_ess}
\end{equation}

\begin{figure}[tb]
\centering
    {\includegraphics[width=0.45\textwidth]{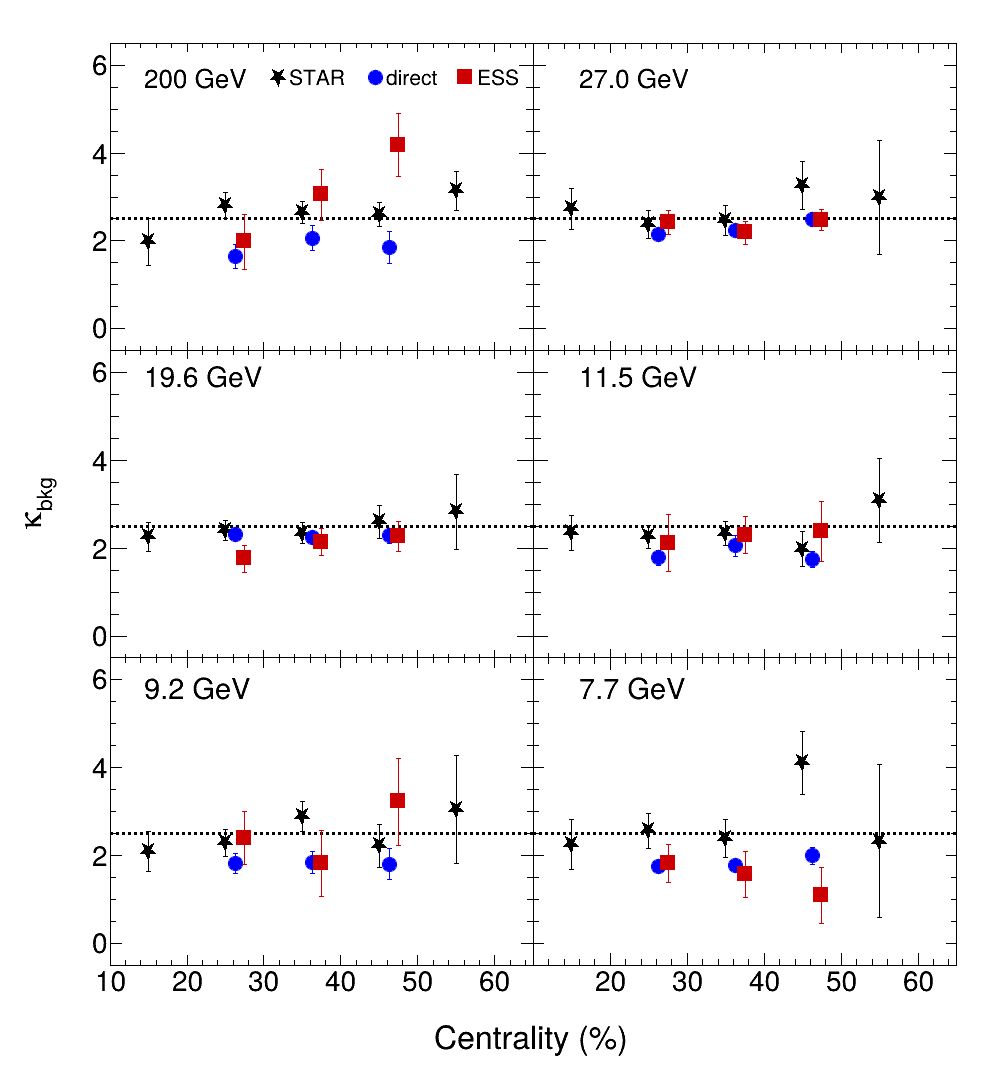}}
\caption{The centrality dependence of $\kappa_{\mathrm{bkg}}$ in Au + Au collisions at various beam energies. The black stars represent the experimental data measured by STAR using the ESS method. The blue circles and red squares denote the AMPT results obtained from direct calculation and the ESS method, respectively. The dashed line indicates $\kappa_{\mathrm{bkg}} = 2.5$.}
\label{fig:k_bkg}
\end{figure}

Fig.~\ref{fig:k_bkg} shows the centrality dependence of $\kappa_{\mathrm{bg}}$ at $\sqrt{s_{\mathrm{NN}}} = 7.7-200$~GeV from the AMPT simulation compared with the STAR results obtained via the ESS method, where $v_{2}\{\mathrm{SP}\}$ is adopted in the calculation.
The values of $\kappa_{\mathrm{bg}}\{\mathrm{direct}\}$ and $\kappa_{\mathrm{bg}}\{\mathrm{ESS}\}$ are consistent with the STAR results within uncertainties, apart from a few data points. 
The $\kappa_{\mathrm{bg}}\{\mathrm{direct}\}$ shows no significant centrality dependence and exhibits a slight increase with increasing collision energy. 
In contrast, due to the large fluctuations of $\Delta\gamma_{\mathrm{ESS}}$, the uncertainties of $\kappa_{\mathrm{bg}}\{\mathrm{ESS}\}$ are too large to draw a firm conclusion regarding its centrality or energy dependence. 
The small value of $\kappa_{\mathrm{bg}}$ observed at 7.7~GeV suggests that certain background contributions, such as those arising from local charge conservation or non-flow correlations, are absent or underestimated in the AMPT simulation at the lowest collision energy.


\section{SUMMARY}

This study investigates the chiral magnetic effect (CME) in Au+Au collisions at $\sqrt{s_{\mathrm{NN}}} = 7.7-200$~GeV using a multiphase transport (AMPT) model with a chiral anomaly transport module. 
The CME signal is extracted via two independent methods: a direct subtraction of the charge-dependent correlator $\langle N_{\text{part}}\Delta\gamma \rangle$ between simulations with zero and non-zero chiral chemical potential $\mu_5$, and the event-shape-selection (ESS) approach, which exploits the linear dependence of $\Delta\gamma$ on the elliptic flow $v_2$.

The background contributions, characterized by the parameter $\kappa_{\mathrm{bg}}$, are evaluated using both the direct and the ESS methods. 
The direct background coefficient $\kappa_{\mathrm{bg}}\{\mathrm{direct}\}$ shows no significant centrality dependence and increases slightly with collision energy
The small background at 7.7~GeV suggests that certain non-flow correlations are not fully captured by AMPT at the lowest energy.

A significant CME signal is observed in the energy range $\sqrt{s_{\mathrm{NN}}} = 9.2-27$~GeV, where the AMPT results with the chiral anomaly provide a reasonable description of the STAR experimental data. 
The extracted CME fractions from the direct and ESS methods are consistent with each other within uncertainties across all energies except at 7.7~GeV, where the ESS method yields an anomalously large CME fraction, likely due to the breakdown of the ESS assumptions in the low-energy regime. At the 200~GeV, the CME signal is strongly suppressed by small chiral anomaly, the strong hadronic phase interaction and the rapid decay of the magnetic field. 
At 7.7~GeV, the CME signal vanishes because the short-lived partonic phase and weak initial magnetic field.

Overall, the AMPT results are in qualitative agreement with the STAR measurements and the ESS-based CME limits. 
The findings highlight that the observability of the CME depends not only on the presence of a chiral imbalance but also on the lifetime of the magnetic field and the duration of the partonic phase, both of which vary strongly with collision energy.

\section{ACKNOWLEDGMENTS}
This work is supported in part by the National Natural Science Foundation of China (NSFC) Grant Nos. 12235016, 12221005, 12305146, 12325507, 12547102 and 12147101, and the National Key Research and Development Program of China under Grant No. 2022YFA1604900, the CAS Project for Young Scientists in Basic Research under Grant No. YSBR-088, and the Guangdong Major Project of Basic and Applied Basic Research under Grant No. 2020B0301030008 and Youth Science Foundation, Grant No. 12205309.

\bibliographystyle{IEEEtran}
\bibliography{reference.bib}
\end{document}